\documentclass[runningheads]{llncs}

\usepackage{amssymb,amsmath}
\usepackage{graphicx,color}
\usepackage[font=small]{subfig,caption}
\usepackage{verbatim} 
\usepackage{hyperref}

\let\doendproof\endproof
\renewcommand\endproof{~\hfill\qed\doendproof}

\spnewtheorem{observation}{Observation}{\bfseries}{\itshape}
\newcommand{\R}{\mathrm{I\!R}} 
\newcommand{\dist}{\text{dist}}

\title{An Interactive Tool to Explore and Improve the Ply Number of Drawings}
\author{Niklas Heinsohn, Michael Kaufmann}
\institute{%
    Wilhelm-Schickhard-Institut f\"ur Informatik, Universit\"at T\"ubingen, Germany\\ \email{\{heinsohn, mk\}@informatik.uni-tuebingen.de}}%

\authorrunning{N. Heinsohn, M. Kaufmann}
\begin{document}

\maketitle

\begin{abstract}
Given a straight-line drawing $\Gamma$ of a graph $G=(V,E)$, for every vertex $v$ the ply disk $D_v$ is defined as a disk centered at $v$ where the radius of the disk is half the length of the longest edge incident to $v$. The ply number of a given drawing is defined as the maximum number of overlapping disks at some point in $\R^2$. Here we present a tool to explore and evaluate the ply number for graphs with instant visual feedback for the user. We evaluate our methods in comparison to an existing ply computation by De Luca et al.~[WALCOM'17]. We are able to reduce the computation time from seconds to milliseconds for given drawings and thereby contribute to further research on the ply topic by providing an efficient tool to examine graphs extensively by user interaction as well as some automatic features to reduce the ply number.
\end{abstract}

\section{Introduction}
Graphs are the common mathematical model to represent relationships between objects and occur in a huge variety of applications and disciplines. 
To make the data stored in a graph accessible for humans, we need a graphical representation which usually involves a drawing of the underlying graph.
There exist several schemes to draw graphs \cite{DBLP:books/ph/BattistaETT99,DBLP:reference/cg/TamassiaL04}. 
In this work we will focus on straight-line drawings.
Several aesthetic criteria on straight-line drawings have been defined to capture the user requirement for a better understanding of the data (eg. edge crossings or angular resolution \cite{DBLP:books/lib/Atallah99}).

Recently a new parameter called ply number was introduced as a quality metric for graph layouts \cite{DBLP:conf/iisa/GiacomoDH0KLMSY15}.
Given a straight-line drawing $\Gamma$ of a graph $G=(V,E)$, for every vertex $v$ the ply disk $D_v$ is defined as a disk centered at $v$, where the radius of the disk is half the length of the longest edge incident to $v$. The ply number of $\Gamma$ is the maximal number of overlapping disks at any point in $\R^2$. Theoretical results have been obtained on the ply number of graphs \cite{DBLP:conf/gd/AngeliniBBH0KSV16,DBLP:conf/iisa/GiacomoDH0KLMSY15} and there exist many real world graphs admitting natural drawings with low ply number \cite{DBLP:conf/gis/EppsteinG08}. 
One common approach to draw such graphs are force-directed algorithms \cite{DBLP:reference/crc/Kobourov13}, whose drawings are known to be aesthetically pleasing. 
A recent study evaluated the correlation between the ply number of drawings produced by force-directed algorithms and other known metrics defined on these algorithms~\cite{DBLP:conf/walcom/LucaGDKL17}. 

There exist many tools and layout algorithms for graph drawing provided e.g. by OGDF \cite{DBLP:reference/crc/ChimaniGJKKM13} or the yFiles library \cite{DBLP:books/sp/04/WieseE004}.
The identification of properties as well as the development of new strategies to optimize parameters of drawings involve frequent examination of graph drawings. 
In this paper we present a tool which allows investigation of graphs according to its ply number. We present a fast algorithm to compute the ply number for a given drawing based on a plane-sweep algorithm which is known to be a powerful technique in computational geometry. Furthermore we provide methods to modify the drawing to reduce the ply number interactively as well as automatically. 
We confirm the value of our tool by providing experimental results on the computation in terms of accuracy and speed as well as the results on our ply minimization approaches. 

\begin{figure}[t]
	\centering
	\subfloat[\label{Fig:Random}]{
		\includegraphics[width=.23\linewidth]{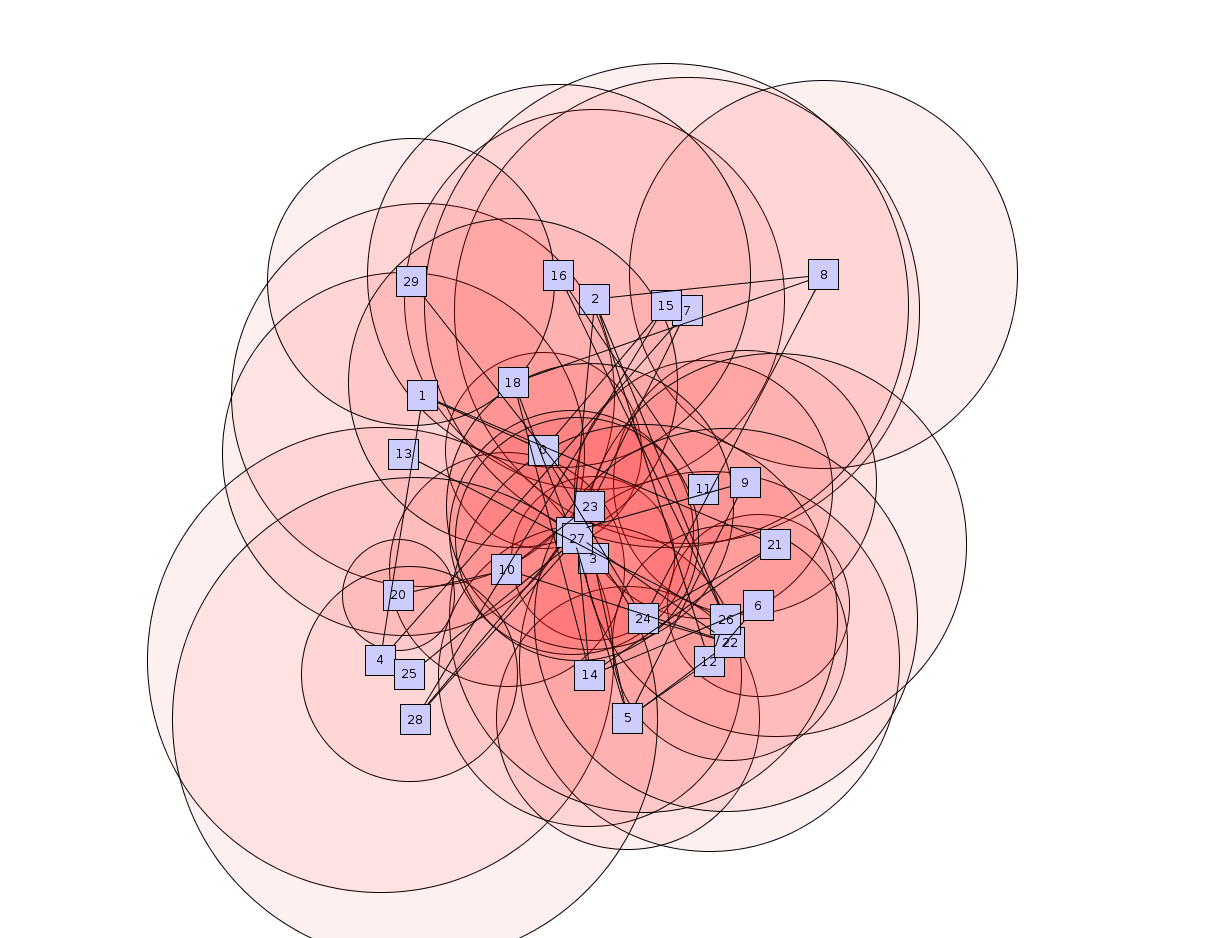}
	}\hfil
	\subfloat[\label{Fig:Circular}]{
		\includegraphics[width=.23\linewidth]{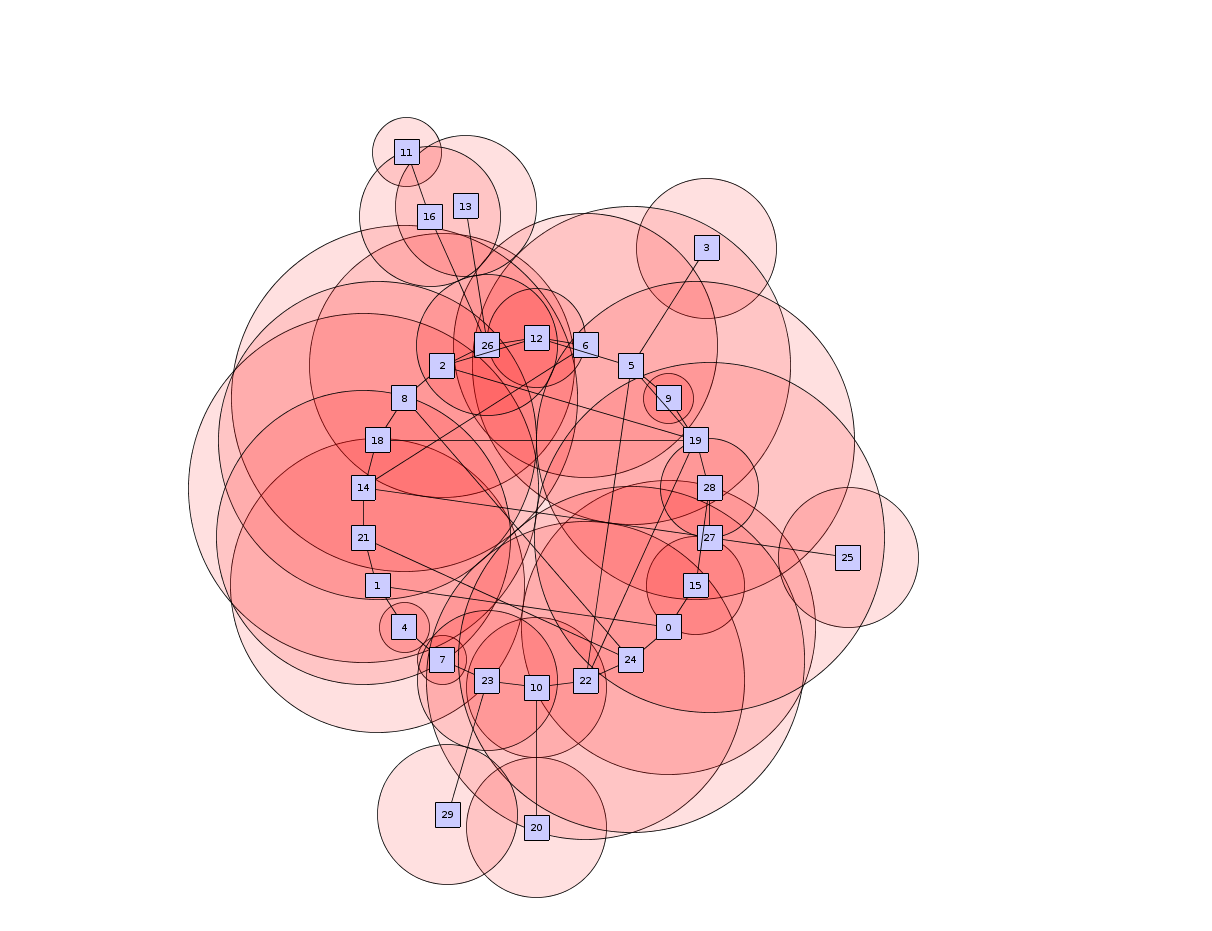}
	}\hfil
	\subfloat[\label{Fig:Organic}]{
		\includegraphics[width=.23\linewidth]{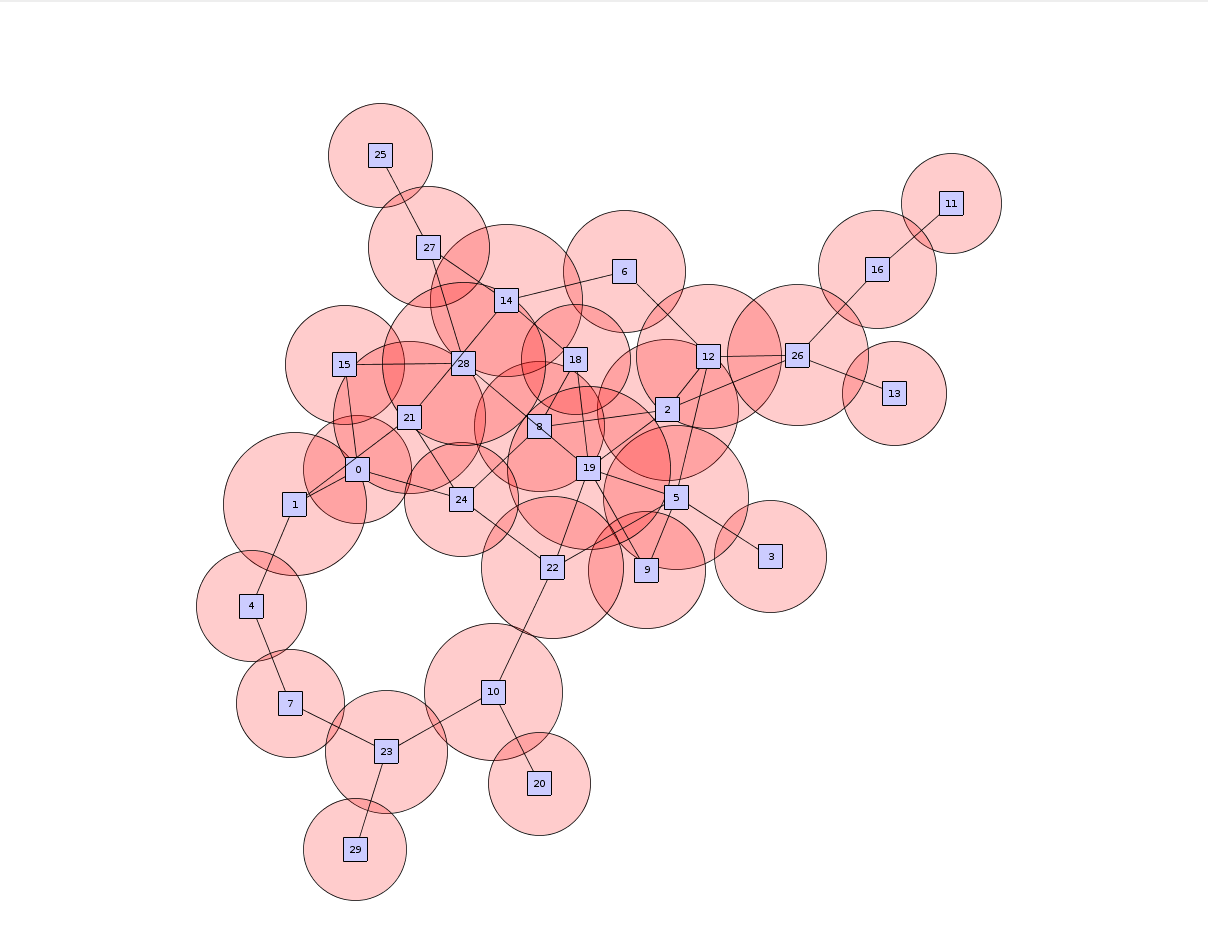}
	}\hfil
	\subfloat[\label{Fig:Modified}]{
		\includegraphics[width=.23\linewidth]{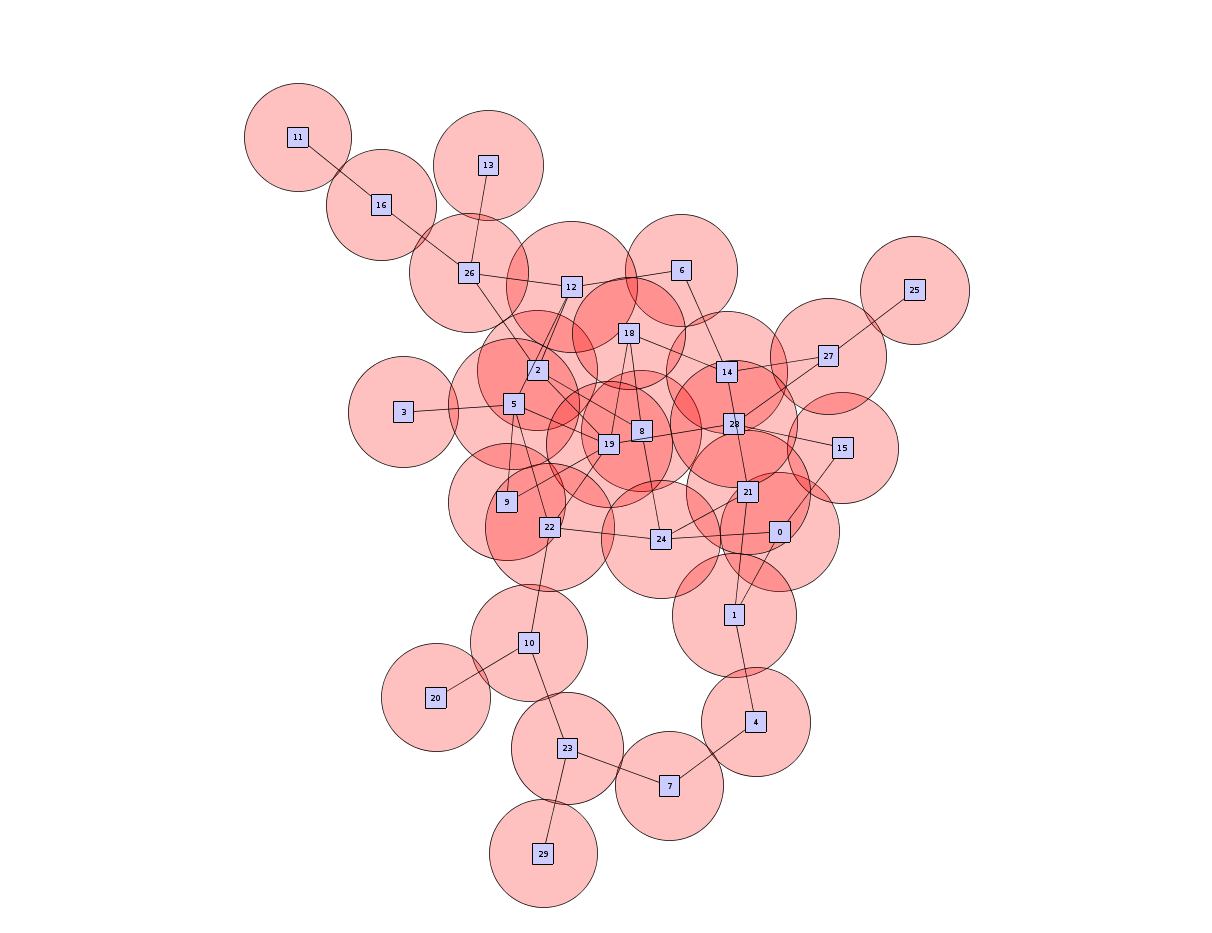}
	}
	\caption{
	A graph with 30 vertices and 40 edges is drawn by our tool in different layouts: (a) randomly placed vertices with ply 15, (b) circular layout with ply 7, (c) organic layout with ply 4, (d) the lowest known ply number of 3 for this graph.}
	\label{fig:intro}
\end{figure}

A first prototype of the basic plane-sweep algorithm has been implemented in~\cite{BachusThesis}.
We reimplemented the algorithm, improved it and added new features.
The experimental study of De Luca et al. \cite{DBLP:conf/walcom/LucaGDKL17} set a benchmark on the computation of the ply number for a given drawing. The authors evaluated several layouts by force directed algorithms according to the ply number. To make our comparison possible, the authors kindly provided some of their data. 
Both implementations of the ply computations are based on the Apfloat library \cite{tommila2003c++} which allows calculations on high precision levels at the cost of time. 

\section{Functionality}
Our tool allows the investigation of a graph regarding its ply number.
As basic functionality, the tool is equipped with some graph layout algorithms, namely organic, circular and randomized (provided by yFiles library \cite{DBLP:books/sp/04/WieseE004}), as presented in \mbox{Figure \ref{fig:intro},} and allows for interactive manipulation of the drawing, such as moving vertices.
Basic file formats are graphml \cite{DBLP:reference/crc/BrandesELP13} and gml \cite{himsolt1997gml} where graphml provides structural information on the graph and gml provides a drawing.

Furthermore, we provide a test if a given drawing is empty-ply, where no vertex is contained in any other vertex's disk. With our tool we identified that the complete graph $K_{4,6}$ admits an empty-ply drawing whereas this was previously known only for $K_{4,4}$ \cite{emptyplysubmission}.
Our implementation is able to compute the ply number for the drawing during runtime, meaning while the drawing is modified, for example by layout algorithms or user interaction.

Live feedback of the ply number on interactive graph manipulations by users is another feature of our tool.
We mark regions where the maximal ply number occurred.
The user can choose between different layouts as start configurations for the graph and is able to improve the ply values automatically by a spring embedder or by manually manipulating the positions of the vertices accordingly.

\section{Ply Computation Algorithm}
To compute the ply number of a drawing $\Gamma$ we have to compute the intersections of the ply disks.
Clearly the ply number is at most linear in the number of vertices. 
In the following, we introduce one major issue in computing ply numbers and present our fast plane-sweep algorithm to compute the ply number. 

\subsection{Precision Problems}
Naturally thinking of an easy case to start with are graphs that admit a drawing with a ply number of 1. This case is easy to describe and points out the difficulty of computing the ply number:
A graph that admits a drawing with ply number 1 has no overlapping ply disks and
can be drawn such that every edge has equal length $l$ and any two vertices are at a distance of at least $l$. 
Suppose that there exists a drawing $\Gamma$ and a vertex $v$ with different edge lengths $|(u,v)| = l_1$ and $|(w,v)|= l_2$ and let w.l.o.g $l_1 > l_2$. $\frac{l_2}{2}$ is a lower bound on the ply disk $D_w$ and since $\frac{l_1}{2}+ \frac{l_2}{2} > l_2$ the ply disks $D_v$ and $D_w$ intersect and the ply number of $\Gamma$ is $\geq 2$. Furthermore, since the radius for any ply disk $D_v$ is $\frac{l}{2}$, the distance $\dist(v,w)$ between any two vertices has to be $\geq l$.  

The complete graph $K_3$ admits a drawing with ply number 1, since the vertices can be placed on an equilateral triangle. Computing a drawing of $K_3$ with the vertices $u,v,$ and $w$, some coordinates must be irrational, since otherwise the condition $\dist(u,v) = \dist (u,w) = \dist(v,w)$ is violated. 
As a consequence the computer would need an infinite precision to represent a drawing with ply number 1.
Furthermore the calculation of coordinates for intersection points of circles involves precise arithmetic and is likely to result in irrational coordinates. 

In previous applications \cite{BachusThesis,DBLP:conf/walcom/LucaGDKL17} this problem was tackled by increasing the precision using the Apfloat library. This allows calculation on up to 1000 digit decimal precision. On the downside these arithmetics require high computational effort.
In the following we present an alternative approach using the primitive type \texttt{
double} reducing the computational effort. 
Later, we evaluate this decision by experiments regarding the precision of the outcome in terms of events and time spent computing the ply number.

\subsection{Plane-sweep Algorithm}
A plane-sweep algorithm describes a powerful technique in computational geometry. 
Given a two-dimensional Euclidean space, a conceptual line which represents the actual state of computation sweeps the Euclidean space scanning for events from left to right. 
Given a drawing $\Gamma$ of a graph $G=(V, E)$, we can easily compute the set of ply disks $D = \{D_v | v \in V\}$.
Every disk $D_v$ is associated with the vertex $v$ at position $(x_v, y_v)$ and radius $r_v$.
At every x-coordinate of the setting there exists a highest ply value. Note that the ply number of the graph is the maximum over all ply values.
The ply value can change whenever a disk starts at $(x_v - r_v, y_v)$, ends at $(x_v + r_v, y_v)$ or if there is an intersection of two disks.
For our purpose these coordinates are called events.

To describe the vertical structure with opening and closing disks, we represent the disks as bottom and top halfcircles, respective to the center. At any x-coordinate between two consecutive events the order of opening and closing halfcircles on a vertical line in a ply drawing and thereby the maximal ply value is fix. For every halfcircle we associate and store the ply value.

Whenever we meet a leftmost coordinate $x_v - r_v$ of a disk $D_v$, we introduce two halfcircles ($h^t_v$ and $h^b_v$) and add them in the vertical structure.
We set the ply values according to the neighboured halfcircles in this case.
Whenever we meet a rightmost coordinate $x_v + r_v$ of a disk $D_v$, we remove both corresponding halfcircles from the vertical structure. If there exist halfcircles between our removed ones, we decrease the ply of these.

Note that two disks might intersect if and only if any two corresponding halfcircles occur next to each other in the vertical structure. Furthermore, any two disks $D_u$ and $D_v$ can intersect at most twice. To keep the computational effort minimal the intersection of disks is calculated the first time any two halfcircles appear next to each other in the vertical structure.
Finally an intersection-event swaps the two affected halfcircles $h_u$ and $h_v$, the ply is updated due to a case distinction.

The events are stored in a priority queue and are executed by their x-coordinate.
In case there exist several events at the same x-coordinate we define the priority in ascending order as end-event, intersection-event, start-event.  

The x-coordinates might get slightly inconsistent due to previously mentioned precision errors.
This results in events which cannot be handled consistently. One example would be an intersection-event that requires a swap of halfcircles, which are not neighboured in the actual state.
Our solution to this scenario is linearly searching for the closest consistent event. This event will be executed and we jump back to the unresolved one. We repeat this until it can be resolved. These events will be tracked as \textbf{postponed events}.
In the results section we evaluate this delay and describe graphs where this periodically occurs.

\section{Experiments}
This section is subdivided into three major parts. At first we will present results on the ply number for various graphs, as well as the number of events and the time. 
Second, we compare our results with the results of De Luca et al.~\cite{DBLP:conf/walcom/LucaGDKL17}. 
Third, we present an approach to reduce the ply number by local modification.

We will make use of three datasets. 
First we take a subset of the Rome graphs \cite{DBLP:journals/comgeo/WelzlBGLTTV97}, which is determined by taking all graphs matching the file name pattern grafo9*. We call the set \textbf{Rome9data}. It consists of 1094 sparse graphs with 10 to 100 nodes and a density between 0.9 an 1.8. 

The second set contains 100 randomly generated graphs. Every graph consists of 100 vertices and has densities between 1.5 and 40. We refer to this set as \textbf{RANDdata}.

The third set will be referred to as \textbf{FM3data}. This set of graphs was kindly provided by the authors of the experimental study \cite{DBLP:conf/walcom/LucaGDKL17} and each graph was drawn using the fast multipole multilevel method of Hachul and J\"unger \cite{DBLP:conf/gd/HachulJ04} which is among the
most effective force-directed algorithms in the literature \cite{DBLP:journals/jgaa/HachulJ07}. 
\textbf{FM3data} contains caterpillars, planar and general (connected simple graphs, generated with uniform probability distribution) graphs. 

\subsection{Ply number for different layouts}
\begin{figure}[t]
\begin{minipage}[b]{.47\textwidth}
\centering
\includegraphics[width=.85\textwidth]{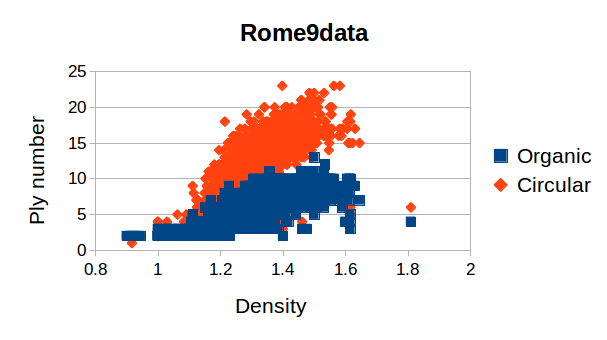}
\captionof{figure}{Ply number for the Rome graphs with various densities.}\label{Fig:Rome9ply}
\end{minipage} \hfill
\begin{minipage}[b]{.47\textwidth}
\centering
\captionof{table}{The table presents the average values on \textbf{Rome9data}.}
\begin{tabular}{l|ccc}
Layout 		& Ply	& Time(ms) 	& Events\\\hline
Organic 		& 6.3 	& $<1$			& 546 \\
Circular 	& 12.5	& 1.1			& 974\\
Random 		& 30		& 3.8			& 3153\\
\end{tabular}
\vspace{.6cm}
 \label{table:SparseLayouts}
\end{minipage}
\end{figure}

Our tool supports different layouts provided by the yFiles library, namely organic, circular and random layout. 
We evaluated our algorithm on \textbf{Rome9data}. We observe that for sparse graphs the organic layout creates drawings with lower ply than the circular layout, whereas at densities close to one, i.e. tree like, they produce similar ply numbers (see Figure \ref{Fig:Rome9ply}). As a reference the random layout produces the highest ply numbers (see Table \ref{table:SparseLayouts}).
We observe a correlation between the ply number, time and events. 

The results on \textbf{RANDdata} are presented in Table \ref{table:randomDensity}. Here, the number of postponed events is noteworthy. 
While it can be neclected in the organic and random layout, the number of postponed events
in the circular layout is consequential and highly increased with the density. 
The highly symmetric placement of the vertices in the circular layout causes many events to share an x-coordinate. 
This circumstance, paired with eventually occurring precision errors, influences this number.
Since summing up over all postponed events exceeds the total number of executed events, the postponed events highly influences the computation time, as we do a linear search for the next event.

\begin{figure}[t]
\centering
\includegraphics[width=.7\textwidth]{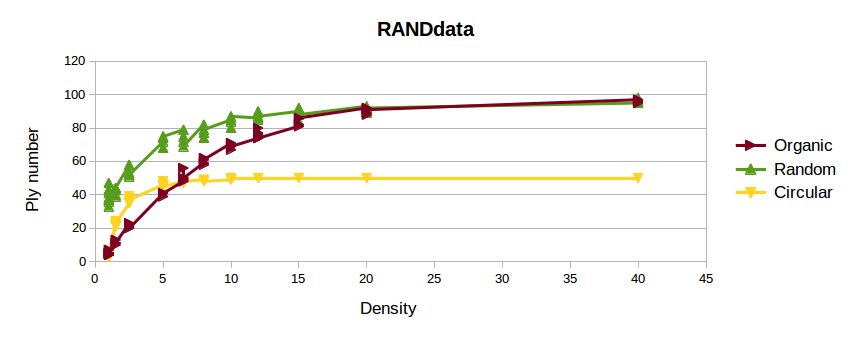}
\captionof{figure}{The density of the graphs of \textbf{RANDdata} is plotted against the ply number of the drawing. We observe that the organic layout produces low ply numbers for low densities, whereas at higher densities the circular layout outperforms the organic layout. In very dense graphs the organic layout performs evenly as the random layout.}\label{Fig:randomDensity}
\end{figure}

In sparse graphs spring embedding algorithms like the organic layout algorithm produce drawings with low ply numbers, while in dense graphs the drawings have similar ply numbers to a drawing where the vertices are placed randomly. In the dense graphs the circular layout hits the theoretical upper bound of $\frac{|V|}{2}$ as shown in Figure \ref{Fig:randomDensity}. 

\begin{figure}[t]
\centering 
\captionof{table}{The average ply numbers for each layout is presented for \textbf{RANDdata}, as well as the average computation time and the average number of postponed events.}
\begin{tabular}{cl |cccc}
Density & Layout 		& Ply 	& Time(ms) 	& Events & postponed\\\hline
&Organic 		& 16.2 	& 2.2			& 2381 		& 0 \\
1.5 - 2.5 &Circular 	& 29.3	& 5.3			& 4349.5 	& 11.3\\
&Random 		& 48.2	& 8.3			& 7170 		& 0\\\hline
&Organic 		& 50 	& 12.2			& 7568 		& 0 \\
5 - 8&Circular 	& 47.5	& 15.9			& 9076.5 	& 109.8\\
&Random 		& 75 	& 15.2			& 9510 		& 0\\\hline
&Organic 		& 76.5 	& 14				& 9534 		& 0 \\
10 - 15&Circular 	& 49.7	& 18				& 9650.8 	& 3343.6\\
&Random 		& 86 	& 14.4			& 9882.5 		& 0\\\hline
&Organic 		& 93 	& 15.9					& 10016.4 		& 0 \\
20-40&Circular 	& 50 	& 160.1				& 9925.6 	& 151232 \\
&Random 		& 94 	& 17.9			& 10041.4 		& 0\\
\end{tabular}
\label{table:randomDensity}
\end{figure}

\subsection{Comparison on the FM3 drawing dataset}
We compare the number of events and the average computation time on \textbf{FM3data}. The Apfloat decimal precision was set to 20 digits for this comparison. This value was used in the experiments of \cite{DBLP:conf/walcom/LucaGDKL17}.
We will present the data split by the type of graphs and according to their density. 

\textbf{FM3data} contains 50 caterpillars with 250 to 450 vertices. The average number of events and the computation time is presented in Table \ref{Tab:Caterpillar}. We observe a huge difference in the total number of events and the computation time. The difference in number of events can be explained due to the difference in handling inconsistent events. The algorithm of \cite{DBLP:conf/walcom/LucaGDKL17} introduces a number of redundant events to detect and handle inconsistencies. Our algorithm reduces the computation for the ply number from seconds to milliseconds.
\begin{figure}[t]
\centering 
\captionof{table}{The caterpillars of \textbf{FM3data}. Each subset with 250 to 450 vertices contains 10 graphs. During all experiments the ply numbers for both implementations were the same. The table presents average values for each set of graphs.}\label{Tab:Caterpillar}
\begin{tabular}{ cc|cc|cc}
& & \multicolumn{2}{c}{\textbf{\cite{DBLP:conf/walcom/LucaGDKL17}}}&\multicolumn{2}{|c}{\textbf{Our Tool}}\\
  Vertices & Ply & Events & Time(ms) &  Events & Time(ms)\\\hline
  250 & 3.8 & 2692.4 & 1328.1 & 1122.6 & 3.5\\
  300 & 4.3 & 3510.4 & 1831.9 & 1430 & 1.6\\		
  350 & 4.5 & 3827   & 1883.7 & 1602.4 & 1.9\\		
  400 & 4.6 & 4564.9 & 2291.5 & 1879.3 & 2.4\\			
  450 & 4.3 & 5032.8 & 2581   & 2110 & 2.3
\end{tabular} 
\end{figure}

\begin{figure}[b]
\centering
\includegraphics[width=.5\textwidth]{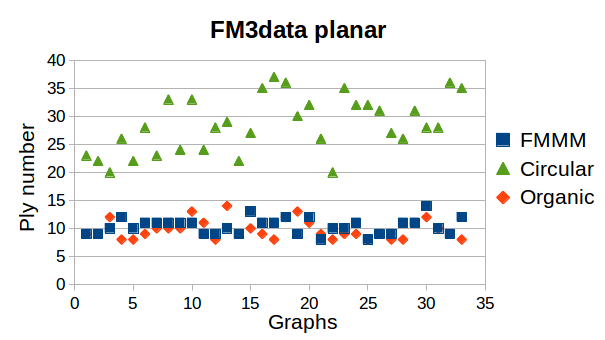}
\captionof{figure}{For each planar graph of \textbf{FM3data} the computed ply numbers for each of the three layouts is plotted. Note that the organic and the FM3 drawings have similar ply numbers, while the circular layout produces higher ply numbers on low density graphs.}\label{Fig:FM3planar} 
\end{figure}

\textbf{FM3data} also contains planar graphs ranging from 250 to 400 vertices and a density ranging from 1.5 to 2. We split the planar graphs in two density classes. In one set all densities are $\leq 1.7$ and in the second set the densities are $> 1.7$. The results are presented in Table \ref{Tab:FM3planar}.
We observed that the ply number seems to be related to the density rather than the number of vertices. A summary is presented in Figure \ref{Fig:FM3planar}.
As a confirmation of previously made observations we computed the ply of organic and circular layouts. For low densities the circular layout produces higher ply drawings where the organic layout produces similar ply numbers as the FM3 algorithm. 
On average the drawings generated by FM3 have slightly higher ply than the organic layout. The average ply number in the organic layout is 9 for graphs with density $\leq 1.7$ and 9.7 for higher density.
Again, note the difference in number of events and computation time keeping results equal.
\begin{figure}[t]
\centering
\captionof{table}{The planar graphs of \textbf{FM3data}. The values show the average results for each subset. Both implementations always computed the same ply numbers.}\label{Tab:FM3planar}
\begin{tabular}{l c|cc|cc}
& & \multicolumn{2}{c}{\textbf{\cite{DBLP:conf/walcom/LucaGDKL17}}}&\multicolumn{2}{|c}{\textbf{Our Tool}}\\
  Density & Ply & Events & Time(ms) &  Events & Time(ms)\\\hline
  $\leq 1.7$ & 9.6 & 9878.6 & 8444.2 & 3434.6 & 3.7\\
  $> 1.7$  & 11.4 & 9625.9 & 9625.9  & 3609.4 & 3.8 \\
\end{tabular}
\end{figure}

The remaining subset of \textbf{FM3data} consists of general graphs with 250 to 450 vertices and the densities 1.5 and 2.5.
In 96 of 100 graphs both \mbox{implementations} computed the same ply number, whereas in 3 graphs we see a difference of 1.
In one specific graph, namely \texttt{General\_400\_2.5\_5\_d\_0\_FMMM\_drawing.gml}, the ply number differs by~5. In these graphs we can detect a high number of postponed events.
Furthermore, our algorithm underestimates the ply number in all cases.
\begin{figure}[t]
\centering 
\captionof{table}{The average results of the general graphs of \textbf{FM3data} are presented. The ply numbers in brackets indicate a different result of the algorithms. Note that these cases have a high number of postponed events. }\label{Tab:FM3results}
\begin{tabular}{c cc|cc|ccc}
 & & & \multicolumn{2}{c}{\textbf{\cite{DBLP:conf/walcom/LucaGDKL17}}}&\multicolumn{3}{|c}{\textbf{Our Tool}}\\
Density & Vertices & Ply & Events & Time(ms) &  Events & Time(ms) & postponed\\\hline
 & 250 & 18 & 18334 & 23955 & 6430 & 10.2 & 0.4 \\
 & 300 & 19.8 & 25688.3 & 38454.8 & 8950.8 & 9.7& 159 \\
1.5 & 350 & 23.7  & 34140.5 & 52829.1 & 11949.7 & 23.7 & 0.6\\	
 & 400 &25.4
  & 43543.4
 & 72928.5
 & 15227.5
 &16.4
 & 0.3\\
 & 450 & 28 (27.9) & 55643.2
 &100653.2
  & 19395.6
 & 23.1
& 0.7\\ \hline
  & 250 & 38.1 & 47248 & 92192.7 & 16539.1 & 19.5 & 0.5 \\
 & 300 & 45.4  & 68070.5 & 147113.6 & 23892.3 & 36.7 & 2.2 \\	
2.5 & 350 & 51.4 & 90943.4
 & 217999.9
 & 31850.1
 & 43.5& 2.5\\
 & 400 & 59.3 (58.7)
 & 118188.7
 & 309601.8
 & 40606.9
 & 83.8
& 40048.3
\\
 & 450 &64.3
(64.2)
 &148973.3
  &426993.5
 & 51640.4
 & 112.9
&62776.7
\\
\end{tabular} 
\end{figure}

\begin{figure}[t]
\includegraphics[width=\textwidth]{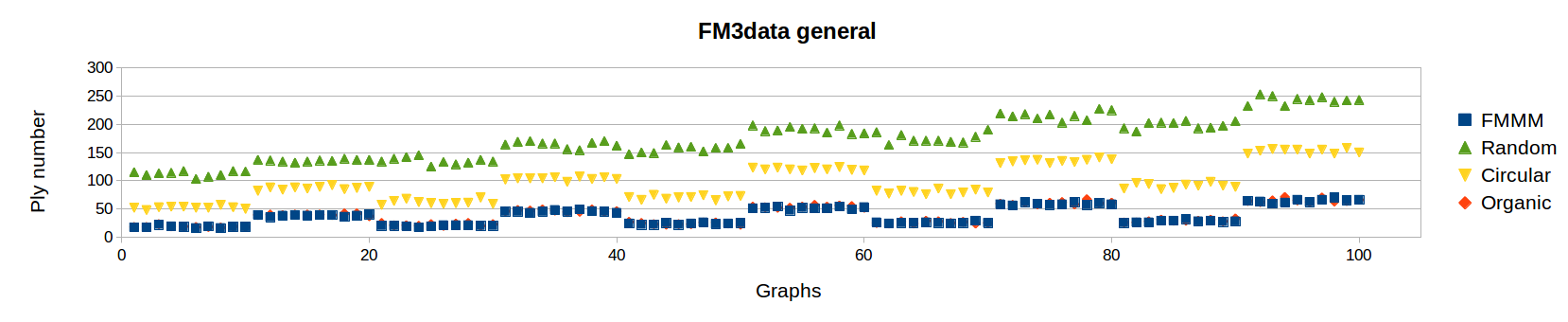}
\caption{The set of general graphs can be subdivided in 5 subsets consisting of graphs with 200, 250, ... , 450 vertices. Each subset can be divided in 10 graphs with density 1.5 and 10 graphs with density 2.5. The figure indicates that the \textbf{FM3} algorithm and the organic layout produce similar drawings regarding the ply number. }\label{Fig:FMMMGeneral}
\end{figure}

To conclude this section we present the interesting result on different layouts on the third subset of \textbf{FM3data} presented in Figure \ref{Fig:FMMMGeneral}. 
Note that the ply numbers on FM3 and organic layout are very similar. The stairs in the plot indicate the jump between the densities from 1.5 to 2.5 for each set of graphs.

\subsection{Ply Minimization} \label{Sec:minimi}
In this part we will present some strategies to create drawings with low ply.
We evaluated our strategies on \textbf{FM3data} and \textbf{Rome9data}. 
Our first strategy is based on an obvious upper bound of $\frac{|V|}{2}$ on the ply number of any graph $G=(V,E)$, 
which is obtained by placing the vertices regularly on a circle $C$. 
For every disk there exist a unique disk on the opposite side of the center which is not overlapping. Therefore at most half of the disks can contribute to the ply number. We use this observation and apply the circular layout, whenever the actual layout has ply number larger than $\frac{|V|}{2}$.

\subsection{Strategies}
To achieve a low ply drawing for a given graph we introduce a workflow, which is directly accessible in our tool.
We start with the organic layout since it has presented itself to produce drawings with low ply number on sparse graphs. Examining these drawings with our tool, we observed that there often exist very few regions with the maximal ply 
number, which often can be reduced by moving a few vertices locally.
From these observations we adjusted a new spring embedder based on Fruchterman and Reingold \cite{DBLP:journals/spe/FruchtermanR91}, similar as suggested in \cite{DBLP:conf/walcom/LucaGDKL17}, and tuned the parameters to produce drawings with less ply.

\subsection{Results}
\begin{figure}[t]
\includegraphics[width=\textwidth]{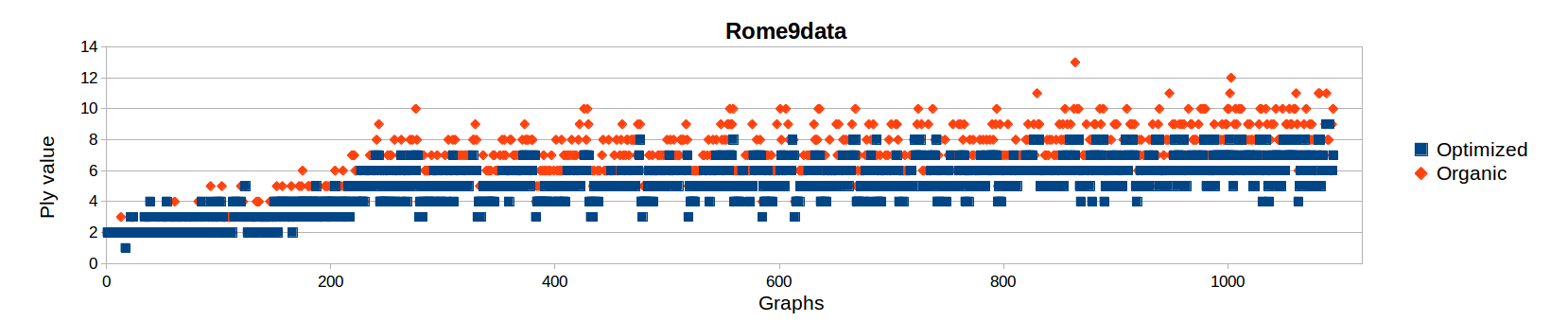}
\caption{For each graph in the \textbf{Rome9data} set the organic and the improved ply number are drawn. The graphs are ordered by the number of vertices. }\label{Fig:Rome9Opt}
\end{figure}

At first we present the advantages of our methods on \textbf{Rome9data} in comparison to organic layouts. On average the ply number of 6.3 of the organic layout was improved to 5.1 in the modified setting. This result is presented in Figure \ref{Fig:Rome9Opt}.

In the experimental study of the ply number \cite{DBLP:conf/walcom/LucaGDKL17} one of the results was the strength of the FM3 algorithm to produce low ply drawings. 
We compare the ply numbers for \textbf{FM3data} to the findings of our ply minimization workflow. Like in the previous chapters we will present the results separately. For all computations of the ply numbers we took our implementation for consistency. 

\begin{figure}[t]
\centering
	\subfloat[\label{Fig:fm3optcat}]{
		\includegraphics[width=.38\linewidth]{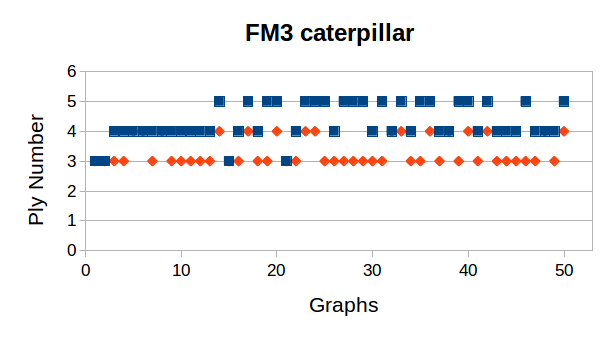}
	}\hfil
	\subfloat[\label{Fig:fm3optplan}]{
		\includegraphics[width=.38\linewidth]{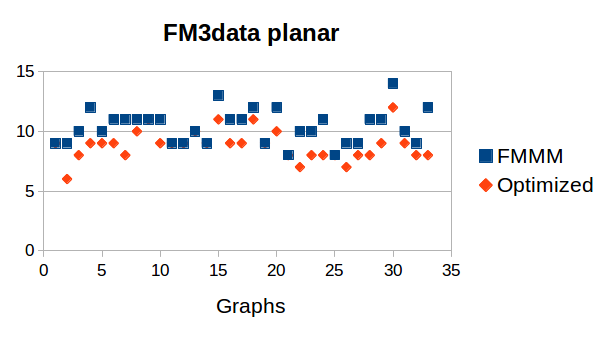}
	}\hfil
	\subfloat[\label{Fig:fm3optgen}]{
		\includegraphics[width=\textwidth]{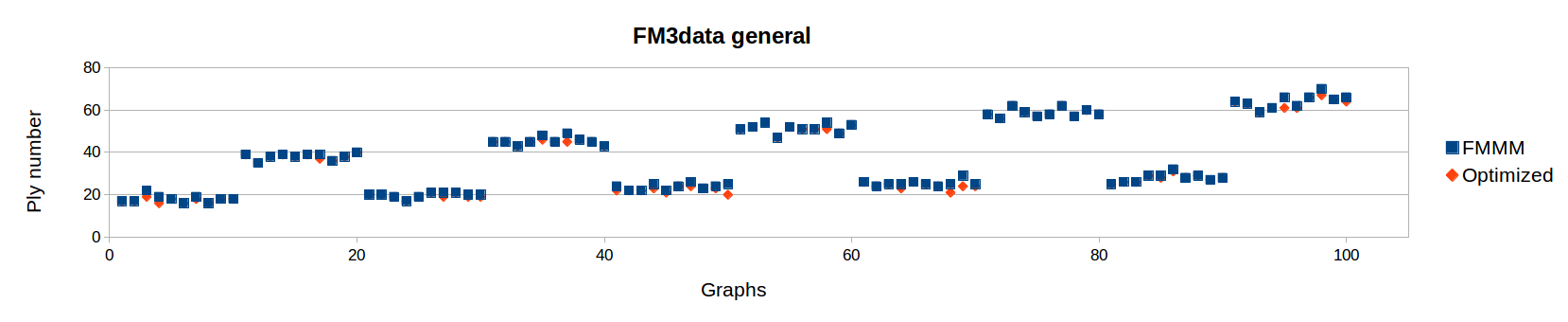}
	}\hfil 
	\caption{The plots present the minimization results on \textbf{FM3data}. Note that the axis change in scale throughout the plots. (a) the ply number of the caterpillars. (b) the ply number of the planar graphs. (c) the ply number of the generel graphs.}\label{Fig:fm3optimization}
\end{figure}

On the caterpillars the average ply number for the \textbf{FM3data} is 4.3, which we could improve to 3.3. On the general graphs we could reduce the average ply number from 37.3 to 36.7 and on the planar subset we could even improve the average ply number from 10.4 to 8.8. 
The results are presented in Figure~\ref{Fig:fm3optimization}.

\section{Discussion \& Conclusion}
To start our discussion we first want to analyze the results for the ply computation for the different layouts followed by the comparison between the two implementations. We continue with the ply minimization part. 
We conclude with a paragraph on the advantages of our tool. 

We have presented the results of the ply computation on various graph layouts. 
Comparing different layouts, we can easily conclude that on sparse graphs spring embedding algorithms produce low ply drawings. This confirms the findings of~\cite{DBLP:conf/walcom/LucaGDKL17}.
Analyzing denser graphs, these algorithms tend to reach their limits. On very dense graphs (close to complete graphs) they perform similar to random layouts (see Figure \ref{Fig:randomDensity}). To strengthen this claim we included the ply number for randomly drawn graphs in this figure.
An interesting observation suggests that the circular layout produces ply numbers close to $\frac{|V|}{2}$ even in the worst case. The reason for this is stated in Section \ref{Sec:minimi}. 

Our experiments suggest the equilibrium between the circular and the organic layout to be between density of 5 and 6.5. At densities larger than 6.5 the ply numbers for graphs drawn with the circular layout are clearly lower than the ply numbers for the organic layout (see Figure \ref{Fig:randomDensity}).

The number of total events indicate a similar observation. While the number of events highly correlates with the increasing density,
a higher number of events seems to imply a higher ply number of the drawing.
Accordingly, the number of events in dense graphs support the observation that our organic layout produces similar ply numbers as the random layout (cf. Table \ref{table:randomDensity}). This effect is expected and 
the reason for this is twofold. On the one hand, every vertex has one ply disk which represents the dependency on the number of vertices. On the other hand, more edges tend to induce larger radii and thereby more intersections even though the number of disks stay the same.
The increasing number of events according to the density and number of vertices is observable in both evaluated implementations.%

According to the precision errors we observe a high number of \textbf{postponed events}, especially in the circular layout. This can be explained by the highly symmetric structure of these drawings, which cause many events to share an x-coordinate.
The radii of the ply disks are likely to be irrational numbers and are thus prone for errors. 
Note that the value purely counts the number of events, which could not be solved instantly. There exist events which are counted several times, since they require more steps in between to be solved and we jump back to the first unsolved event.

Comparing the two implementations on the \textbf{FM3data}, we observe three facts. 
First of all the number of total events differ by a factor of $2$ to $3$. This can be explained due to the implementation of \cite{DBLP:conf/walcom/LucaGDKL17} adds additional events to prevent the influence of precision errors.
This way the errors are detected instantly.
Second, the difference in computation time depends partially on the pure number of events but the major time difference can be explained by the arithmetic computation time on Apfloat values in comparison to the primitive type \texttt{double}. 
Since we present a tool for the examination of different graphs we need a fast computation of the ply number to achieve a feedback for the user within milliseconds. Note that the implementation of \cite{DBLP:conf/walcom/LucaGDKL17} this was not applicable.
To give an overview, there were only 4 out of 100 graphs where a difference in the computed ply number could be observed.
In every case our implementation underestimated the ply in comparison to the other implementation~\cite{DBLP:conf/walcom/LucaGDKL17}. The average error is very low as presented in Table~\ref{Tab:FM3results}.

Examining the results we can state a likelihood or quality of the computed result. Where the number of postponed events is one important indicator of occurred precision errors and evenly important a large number of end-events to solve inconsistencies increase the likelihood of miscomputation.
A feature we want to include in future work is to give visual feedback to the user in that case.

During the experiments we detected a few computations with a high number of \textbf{postponed events} during the analysis of the \textbf{FM3data}.
Examining the graphs, we observed that the FM3 algorithm tends to produce drawings with low average edge length and thereby are likely to induce precision errors. In our layout algorithms larger average edge length where possible. This increases the accuracy of our algorithm and explains the computation error on these graphs.
All in all we present an algorithm which can compete in the computed result and is very fast. We conjecture that the accuracy can be even increased by scaling a given graph. Unfortunately, we cannot support this by experimental data, another task that will be tackled in future work.

Now we want to discuss the results of our minimization approach.
We compare the organic layout and our strategy to reduce the ply number on \textbf{Rome9data}. We adjusted a spring embedder to reduce the ply number for a drawing based on our organic layout. One of the important observations on sparse graphs was that the maximal ply number is often reached in very few regions. On this data, in average, we can reduce the ply number by one. 
For further competition on sparse graphs we compare the FM3 layout algorithm, which was the winning strategy in \cite{DBLP:conf/walcom/LucaGDKL17}.
Our approach creates drawings that are on average one ply lower than this algorithm.(cf. Figure~\ref{Fig:fm3optimization}).
Even though our implementation tends to underestimate the ply number on some \textbf{FM3data} graphs. Our modification produces a larger scaling and we conjecture that our approach constructs drawings with lower ply number and the computations are more resistant to precision errors.

For very dense graphs the spring embedding strategies seem to produce drawings which have a ply number close to random layouts. Nevertheless, for dense graphs we can guarantee an upper bound by using the circular layout, which is included in our optimization.

Examining the minimization strategies on caterpillars of \textbf{FM3data}, the ply numbers still range up to 4. Even though we know that caterpillars admit a ply 2 drawing \cite{DBLP:conf/gd/AngeliniBBH0KSV16}. Further examination on these graphs suggests that our methods are often able to construct drawings with ply number 2 given a suitable start configuration and enough time.
Since we gave a strict time limit during the experiments
we did not manage to produce many ply 2 drawings on this set.

Our tool provides the user with our adjusted spring embedder and the possibility to enforce equal edge lengths. The equal edge lengths can be interpreted as a test if the actual embedding admits a ply 1 drawing. During our experiments, due to precision errors, we did not observe ply 1 drawings by automated layout methods.
The enforced equality of edge lengths includes very strong forces and converges if there exists a ply 1 drawing in the current embedding.

The optimization process involves several iterative computational steps using spring embedding algorithms and computation of the ply number in between. By using these steps and adjusting the vertices manually it is possible for the user to reduce the ply number even further by moving few vertices, since due to a previous observation there often exist only few regions with maximal ply. 

We conclude this part with a short summary of functionality and a forecast for our tool. 
We introduced a fast ply computation algorithm which is able to give instant feedback to user interaction, e.g. whenever the drawing of a graph is modified. We were successfully able to reduce the computation time from seconds to milliseconds. Our tool is equipped with basic layout algorithms and simple automated minimization techniques. The tool can be used to get a deeper understanding of several graph classes e.g. according to the question if there exists a lower bound on the ply number.
In the near future we will include an indicator on the accuracy of the computation. In these cases the implementation providing higher precision in the computation might be used as verification. Furthermore, we want to improve the minimization methods.
Further evaluation and experiments will be necessary to observe the influence of scaling to our computations. 

\paragraph{\textbf{Acknowledgements}}
We specially thank the authors of \cite{DBLP:conf/walcom/LucaGDKL17} for providing their implementation and data to compare with ours.
We also thank  Patrizio Angelini, Lukas Bachus, Michael Bekos, and Felice De Luca for helpful discussions.

\bibliographystyle{splncs03}
\bibliography{biblio}
 
\end{document}